\documentclass[pdflatex,sn-mathphys-num]{sn-jnl}

\usepackage{graphicx}%
\usepackage{multirow}%
\usepackage{amsmath,amssymb,amsfonts}%
\usepackage{amsthm}%
\usepackage{mathrsfs}%
\usepackage[title]{appendix}%
\usepackage{xcolor}%
\usepackage{textcomp}%
\usepackage{manyfoot}%
\usepackage{booktabs}%
\usepackage{algorithm}%
\usepackage{algorithmicx}%
\usepackage{algpseudocode}%
\usepackage{listings}%
\graphicspath{{}}

\usepackage[version=4]{mhchem}		 
\usepackage{array}
\usepackage{adjustbox}
\usepackage{makecell}
\usepackage{chemformula} 
\usepackage{amsmath}
\usepackage[utf8x]{inputenc}
\usepackage{subcaption}
\begin{document}

\title[Article Title]{Quantum
Study of Halogen Substituted anti-B18H22 Borane Clusters
for Optoelectronics}


\author[1]{\fnm{Mahmoud} \sur{Deeb}}\email{mahmoud.deeb311@hotmail.com}
\equalcont{These authors contributed equally to this work.}

\author[1,2]{\fnm{Nabil} \sur{Joudieh}}\email{njoudieh@yahoo.fr}
\equalcont{These authors contributed equally to this work.}

\author*[3,4]{\fnm{Nidal} \sur{Chamoun}}\email{chamoun@uni-bonn.de}
\equalcont{These authors contributed equally to this work.}

\author[2]{\fnm{Habib} \sur{Abboud}}\email{Habib-abboud@iust.edu.sy}
\equalcont{These authors contributed equally to this work.}

\affil[1]{\orgdiv{Department of Physics}, \orgname{University of Damascus}, \orgaddress{ \state{Damascus}, \country{Syria}}}

\affil[2]{\orgdiv{Faculty of Pharmacy}, \orgname{IUST}, \orgaddress{\street{Daraa Highway}, \city{Ghabagheb},  \state{Damascus}, \country{Syria}}}

\affil[3]{\orgdiv{}\orgname{Arabic Language Academy of Damascus}, \orgaddress{\street{P.O. Box 327, Bld 6, AbdulMunim Riadh St.}, \city{Malki} \postcode{} \state{Damascus}, \country{Syria}}}

\affil[4]{\orgdiv{CASP}, \orgname{Antioch Syrian University}, \orgaddress{\street{} \city{Maaret Saidnaya}, \postcode{} \state{Damascus}, \country{Syria}}}


\abstract{We offer a quantum chemical analysis of mono-halogenated borane molecules using DFT and TD-DFT theories, applying the PBE0/def2-SVPD and B3LYP/6-311+G(d) methods as implemented in ORCA, and explore how solvent effects influence electronic transition properties. 
The comparable benchmarks are the archetype anti-\ce{B18H22} denoted as (1) against hypothetical halogenated derivatives: 
7-F-anti-\ce{B18H21} (2), 4-F-anti-\ce{B18H21} (3), and the recently synthesized 4-Br-anti-\ce{B18H21} (4). The analysis includes an optimization of the ground and first singlet excited states, vibrational frequency analysis, 
and a comprehensive spectroscopic profile covering IR, Raman, UV-Vis absorption, and emission spectra. 

The IR spectra of the fluorinated compounds feature a characteristic B-F stretching peak, 
while the Raman spectra closely resemble the parent molecule. 
UV-Vis spectral analysis shows a redshift and oscillator strength enhancement for F at position B7, 
indicating altered electronic properties due to substitution with lighter halogen. 
Furthermore, solvent effects enhance the probability of electronic transitions. Halogene presence led to a decrease of the energy gap EG(LUMO-HOMO) due to the stabilization of LUMO, which implied a redshift in the emission/absorption wavelength spectra, with the largest EG change at around 14\% occurring for the (4)$^{th}$ benchmark compound..
Notably, all compounds emit light within the visible spectrum, underscoring their potential for optoelectronic applications.}





\maketitle

\section{Introduction}\label{sec1}

Boron compounds have recently attracted considerable attention due to their wide-ranging applications in various fields. 
Among the binary boranes, which consist primarily of boron and hydrogen, 
the anti-\ce{B18H22} isomer is particularly notable for its unique fluorescence properties and a quantum yield nearing unity. 
The photophysical characteristics of this isomer were first uncovered by Londesborough and his team in an experimental study 
that was supported by theoretical computational analysis of its excited states \cite{anti2012}. 

The first derivative of this isomer, anti-\ce{B18H20(HS)2}, was synthesized through chemical substitution. 
This modification significantly enhanced singlet oxygen production, increasing the quantum yield from 0.0008 in the original compound 
to 0.59 in the substituted version \cite{hs2013}.

Additionally, the first inorganic laser based on the anti-\ce{B18H22} molecule was developed, with its solutions displaying pulsed laser emission 
in the blue spectral range and achieving an efficiency of 9.5 \% \cite{b2015l}. 
This breakthrough has paved the way for further research into anti-\ce{B18H22} derivatives aimed at enhancing their photophysical properties. 

Researchers have since synthesized and characterized a new di-substituted derivative, anti-\ce{B18H20(NC5H5)2}. 
In solution, this derivative exhibited fluorescence emission in the range of 650-690 nm, though with a low quantum yield of 0.003. 
However, in the solid state, the fluorescence wavelength experienced a blueshift to 585 nm, 
and the quantum yield of fluorescence significantly increased to 0.15 \cite{py2017,py2018}.

The borane compound was treated with iodine, resulting in the formation of mono- and di-halogenated derivatives: 
7-I-anti-\ce{B18H21} and 4,4'-I$_2$-anti-\ce{B18H20}. 
The mono-iodinated derivative exhibited green phosphorescence at 525 nm with a quantum yield of 0.41. 
In contrast, the di-iodinated derivative showed a wavelength shift to 545 nm and a higher quantum yield of 0.71. 
The experimental absorption spectra of both iodinated compounds were similar to that of anti-\ce{B18H22}, 
with a redshift of approximately 30 nm due to iodine's influence. 
Theoretical analysis using the CASPT2 method aligned well with experimental values for vertical absorption and emission energies \cite{i2019}.
The optical stability of the compound 4,4'-I$_2$-anti-\ce{B18H20} is better than that of  7-I-anti-\ce{B18H21}, whereas the radiative lifetime $\tau_0$ of the latter (27 $\mu$s) largely exceeds the former's (2.4 $\mu$s) due to the di-halogenated derivative containing  two iodine atoms enhancing the spin-orbit coupling, and thus more probable intersystem crossing processes, and hence decreasing the excited state lifetime.       
The quantum yield of singlet oxygen photosensitiazation is higher in the mono(0.52)- than the di(0.36)-iodinated  derivative, whereas both compounds are highly luminescent, exhibiting effective green phosphorescence, where emissions could be quenched via dioxygen with high quantum yield. These good efficiencies in the  photophysical processes make these two compounds promising photosensitizers of dioxygen.   

Bromination of anti-\ce{B18H22} produced the mono-halogenated derivative 4-Br-anti-\ce{B18H21Br}, which exhibited dual emission: 
fluorescence at 410 nm and phosphorescence at 503 nm. 
Computational analysis, utilizing the hybrid functional PBE0 and the DZP basis set with relativistic effects, 
showed good agreement between theoretical calculations and experimental data for both absorption and emission spectra \cite{br2020}. The substituted compound absorbs light at 343 nm with a molar absorptivity less than that of the original one by around 72\%.  The dual emission emerges from a delicate balance between populating the two levels $T_1$ and $S_1$, which is provided for by the substituting Br. Quenching the phosphorescence signal is achieved by raising the Oxygen content in the cyclohexane solution, whereas the fluorescence signal is not affected by Oxygen, which allows for the possibility of using this derivative as a ratiometric oxygen probe.

Recently, alkylated borane derivatives were synthesized, demonstrating high stability, solubility, and unique fluorescence properties. 
These compounds emit blue light in the range of 423-427 nm with a quantum yield of 0.71 to 1. 
Their UV absorption, spanning 283-357 nm, closely mirrors that of the parent compound (anti-\ce{B18H22}), with a redshift of around 10 nm. 
Theoretical analysis using the CASPT2 method was performed to explore the photophysical properties of the anti-\ce{B18H18(CH3)4} and anti-\ce{B18H18(C2H5)4} 
compounds.
The calculated results showed a strong correlation with the experimental values for vertical absorption and emission energies \cite{alk2020}. 

Upon methylation of anti-\ce{B18H22}, an increase in the polyhedral volume was observed, leading to the formation of anti-\ce{B18H8Cl2(CH3)12}. 
This substitution enhanced the absorption and solubility properties of the fluorescent molecule. Compared to anti-\ce{B18H22}, 
there is a redshift of approximately 30 nm in the absorption bands and 20 nm in the emission bands, 
indicating significant changes in the molecule's photophysical properties \cite{swell2020}. 

The size swelling of the substituted Boran substructure is caused by the electronic density decrease in the cluster due to the electron withdrawing effect by the more electronegative Carbon atoms. Absorption coefficient increases by around 20\% compared to that of the original compound, whereas the fluorescence quantum yield lags behind, at 76\%, but still too high compared to other familiar standards in the field.

The compound anti-\ce{B18H20(NC9H7)2}, formed by the addition of isoquinoline to anti-\ce{B18H22}, 
exhibited notable aggregation-induced emission in tetrahydrofuran/water. 
It absorbs light between 370-500 nm and shows a red-shifted absorption peak compared to its precursor. 
Its fluorescence spectrum across various solvents shows a single peak within 616-680 nm, 
with the solvent having minimal effect on the absorption and emission peaks \cite{py2020}. 

In a recent study, a halogenation reaction of the borane compound produced mono-halogenated derivatives, 
specifically 4-I-anti-\ce{B18H21} and 7-Cl-anti-\ce{B18H21}. 
Spectroscopic analysis revealed that the iodinated derivative displayed phosphorescence at a wavelength of 514 nm 
with a quantum yield of 0.16. 
In contrast, the chlorinated derivative exhibited fluorescence emission at a shorter wavelength of 418 nm, 
achieving a quantum yield of 0.80, the highest recorded among all halogenated borane derivatives reported to date \cite{cl2022}.

The compound 4-I-anti-\ce{B18H21} shows an absorption similar to that of the original compound, but with a molar absoptivity less by 35\%, similar to the case of 7-I-anti-\ce{B18H21}. It emits a green phosphorescence at 514 nm with a quantum yield (0.16), less than its 7-I-anti-\ce{B18H21} counterpart (0.41). As to the compound 7-Cl-anti-\ce{B18H21}, it emits light at 344 nm with a molar absorptivity larger by 20\% than that of the original one, whereas its fluorescence occurs at 418 nm with quantum yield (0.80). These data show that the absorption coefficient, the emission wavelength    and the nature of luminescence (phosphorescence vs fluorescence) depend on the substituting atom identity, whereas the quantum yield depends more on the substitution position.   

By treating the original compound by $AlCl_3$ in Tetrachloromethane solutions, one gets various Chlorine substituted derivatives.\cite{cl2023} The compound 3-Cl-anti-\ce{B18H21} (3,3'-Cl$_2$-anti-\ce{B18H20}, 4,4'-Cl$_2$-anti-\ce{B18H20}) absorbs light at 327 (324, 344) nm, with fluorescence at 407 (410, 435) nm, with quantum yield of 0.52 (0.07, 0.17) and fluorescence lifetime of 6.1 (1.2, 1.5) ns. The study showed that the increase of the number of  chlorine atoms implies a decrease in the fluorescence quantum yield as well as in its lifetime. Substitution at the B3, B3' positions reduces the fluorescence efficiency more than the case of B4, B4'. Moreover, substitution at B4 leads to a larger (smaller) redshift (fluorescence quenching) than that corresponding to B3 substitution. Also, substitution at B7 is more stable compared to those of B3 or B4, but it provides a better quantum yield, since the Chlorine atom leads to reducing the excited states lifetime, whence the probability of their absorption of the pumping\/stimulated emission energies decreases, a factor which plays a major role in worsening the performance of Boran laser \cite{unveiling}

Since the light halogen Cl showed the highest quantum yield among all halogenated boranes up to date , and led to molar absorptivity which exceeds the parent compound's one, 
it is logical and important to explore the effects of the lightest halogen F on the photophysical properties of laser borane. 
Furthermore, we aim to complete analysing the halogens substitutions, in that the  Br, I and Cl have been studied recently, with no mention of F and At, so it was natural to study these two elements, which was not done previously. While the calculations of the latter (At) are underway, the analysis of F substitutions forms an integral part  of this work. Past studies did not discuss, neither the IR and Raman spectra, nor the thermodynamical properties of Borane. Thus, our results can be helpful in designing a Borane-derivative apt for future optoelectronics applications requiring certain wavelengths and refined thermodynamical conditions.  

Building on existing research, we have conducted a theoretical investigation using computational chemistry to explore the properties 
that arise when a hydrogen atom in anti-\ce{B18H22} is replaced by a halogen atom. 
The goal of this study is to uncover the structural and photophysical changes that such substitutions can trigger in borane compounds. 
We examined the hypothetical fluorinated molecules 7-F-anti-\ce{B18H21} and 4-F-anti-\ce{B18H21}, 
alongside the parent molecule anti-\ce{B18H22} and the recently synthesized brominated derivative 4-Br-anti-\ce{B18H21}. For this work, we have 
used hyprid functionals which can afford a good result in balance offering a compromise between the accuracy and the computational costs.

Previous studies have shown that sites B4 and B7 are the most favored sites for halogen substitution. According to reference \cite{br2020}, the borane heads B4 and B4' are the most electron-rich and therefore the most likely to undergo electrophilic substitution. Moreover, 
the similarities in reaction conditions for producing iodine- and chlorine-substituted compounds at site B7, such as the absence of a catalyst, suggest that the preference for substitution site can be determined by the intermediate reactive species arising from halogen conditions \cite{cl2022}. Most previous studies were limited to halogenations at sites 4 and 7, and we stick in this work to this choice.

The plan of the paper is as follows. In section 2, we present the computational methods that we used. The results are presented in section 3, including the optimised ground state geometry in subsection 3.1, and the HOMO and LUMO orbitals in subsection 3.2, followed by a study of the vibrational modes of IR and Raman spectroscopies in subsection 3.3, while presenting the TD-DFT calculation in subsection 3.4 covering the vertical emission and absorption energies as well as the adiabatic and 0-0 ones. We end up by concluding remarks in section 4. An appendix states all the vertical absorption wavelengths and corresponding oscillation strengths using different methods.  

\section{Computational Methods}
In this study, quantum chemical calculations were carried out using the ORCA software package (version 5.0.2) \cite{orca}. 
The PBE0 hybrid functional and the def2-SVP basis set were utilized for optimizing molecular geometries and analyzing IR and Raman spectra \cite{pbe0,svp}. 
Optimization of the first excited singlet state (\ce{S1}) was performed with the def2-SVPD basis set. 
To calculate the vertical transition energies for the first ten excited singlet states, TD-DFT calculations were conducted using two methods: 
PBE0/def2-SVPD \cite{svpd} and B3LYP/6-311+G(d) \cite{b3lyp,6-311} in both gas phase and solvent phase (Hexane) with the CPCM model. \cite{cpcm}
Molecular structures were visualized, and atom numbering was adjusted using Chemcraft software to ensure consistency with previous studies \cite{chemcraft}. Moreover, the spectroscopic analysis were processed using Python.

PBE0 method is built on theoretical parameters, which makes it more universal in computation. Furthermore, it proved effective in the past Bromine-substituted compound study \cite{br2020}. On the other hand, B3LYP is built on experimental parameters, and is widely used in quantum computations, and also proved to be successful when applied to calculate the equilibrium geometry of the original compound \cite{anti2012}. The basis def2-SVP was sufficient to compute the equilibrium geometries of the ground state, and it was successful in obtaining valid structures. From the other side, the two bases def2-SVPD and 6-311+G(d) provided, both, similar accuracies in calculation, as they both use diffuse functions, which provides for a more precise description of the excited states. 
Finally, we chose the CPCM model, amidst the ORCA available models including the sophisticated SMD and the obsolete COSMO, as it is a compromise between good accuracy and heavy computational cost. We adopted the Hexane solvent since it was used in the previous study of the original compound, which would make the comparison easier.

.

\section{Results and discussion}
\subsection{Optimized Ground State Geometry:}
\begin{figure}[H]
\caption{Optimized \ce{S0} Geometry with atoms numbering.}
\label{geos0all}
\centering
\begin{subfigure}{.3\linewidth}
\includegraphics[width=\linewidth]{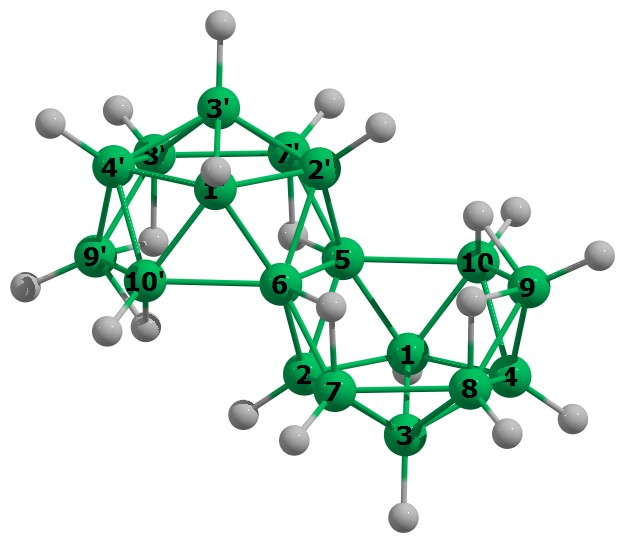}
\caption{anti-\ce{B18H22}}
\label{antis0}
\end{subfigure}\hfill
\begin{subfigure}{.3\linewidth}
\includegraphics[width=\linewidth]{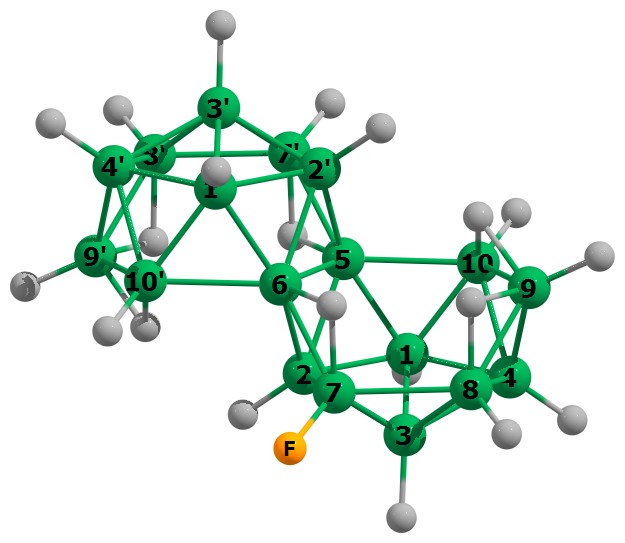}
\caption{7-F-anti-\ce{B18H21}}
\label{7fs0}
\end{subfigure}

\medskip

\begin{subfigure}{.3\linewidth}
\includegraphics[width=\linewidth]{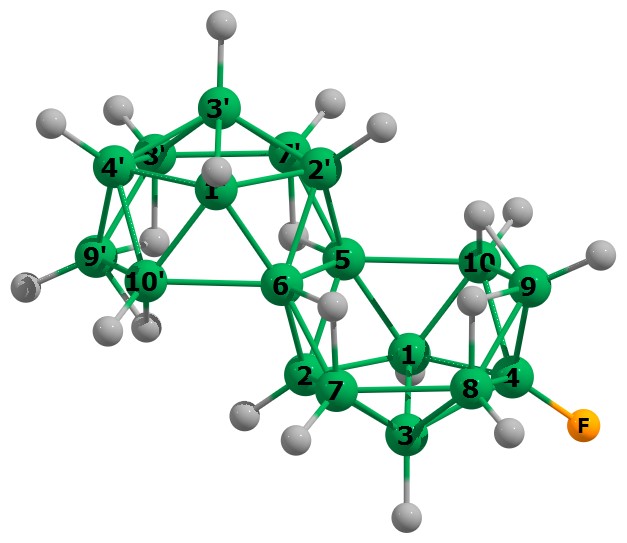}
\caption{4-F-anti-\ce{B18H21}}
\label{4fs0}
\end{subfigure}\hfill
\begin{subfigure}{.3\linewidth}
\includegraphics[width=\linewidth]{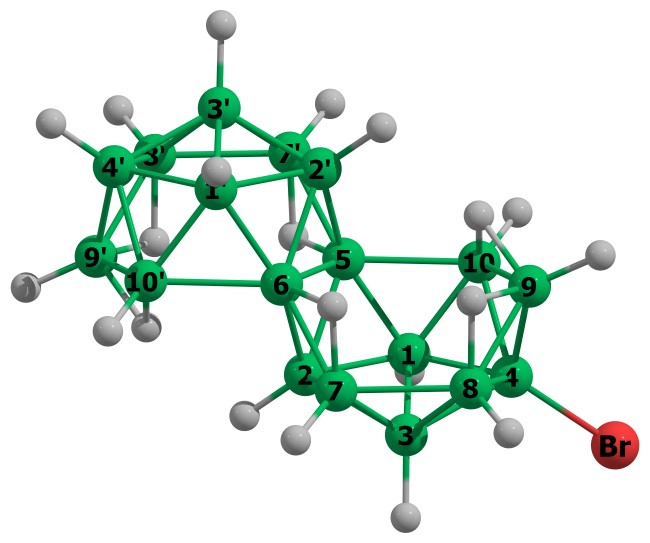}
\caption{4-Br-anti-\ce{B18H21}}
\label{4brs0}
\end{subfigure}
\end{figure}

\begin{figure}[H]
\caption{Maximal differences $\delta$ in B-B bonds between halogenated derivatives and parent compund, at their $S_0$ geometries}
\label{bonds0fig}
\centering
\includegraphics[width=\linewidth]{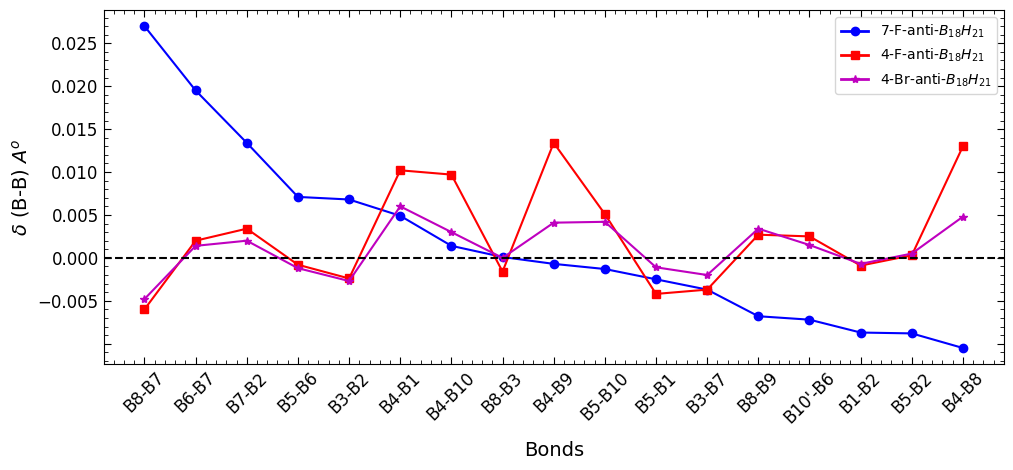}
\end{figure}

The ground-state geometric configuration for each molecule (Figure \ref{geos0all}) was precisely optimized using the PBE0 hybrid functional combined 
with the def2-SVP basis set, as implemented in the ORCA computational chemistry software. 
Generally, the molecular geometry remains mostly unchanged upon substitution, 
with only slight expansions and contractions observed in the boron-boron bond lengths near the substitution sites. 
In the compounds 7-F-anti-\ce{B18H21} (1), 4-F-anti-\ce{B18H21} (2), and 4-Br-anti-\ce{B18H21} (3), 
the most significant bond elongations were found in the B8-B7, B4-B9, and B4-B1 bonds, respectively, 
with deviations from the anti-\ce{B18H22} (1) reference molecule being 0.027, 0.013, and 0.006 Å, respectively. 
Conversely, the most contracted bonds for (2), (3), and (4) were B4-B8, B8-B7, and B8-B7, respectively. 
These bonds showed changes in bond length compared to the parent compound of -0.010, -0.006, and -0.005 Å, respectively, 
highlighting the significant impact of the substitutions at these specific molecular positions. 
Table \ref{bonds0} presents the bond lengths for the studied molecules at their optimized ground-state geometries, whereas figure (\ref{bonds0fig}) shows the maximal deviations $\delta$ in B-B bonds between the halogenated derivatives and the parent compound, at their $S_0$ geometries.

\begin{table}
\caption{Bonds lengths for all derivatives compared to parent molecule}
\label{bonds0}
\centering
\begin{tabular}{lccccccc}
\toprule
\multirow{2}{*}{Bonds} & anti-\ce{B18H22} & \multicolumn{2}{c}{7-F-anti-\ce{B18H21}} & \multicolumn{2}{c}{4-F-anti-\ce{B18H21}} 
& \multicolumn{2}{c}{4-Br-anti-\ce{B18H21}} \\
\cmidrule{2-2} \cmidrule(lr){3-4} \cmidrule(lr){5-6} \cmidrule(lr){7-8}

&d(B-B) $A^o$&d(B-B) $(A^o)$&$\delta$ $(A^o)$&d(B-B) $(A^o)$&$\delta$ $(A^o)$&d(B-B) $(A^o)$&$\delta$ $(A^o)$ \\
\midrule
B8-B7 & 1.9522 & 1.9792 & 0.0270 & 1.9462 & -0.0060 & 1.9474 & -0.0048 \\
B6-B7 & 1.8145 & 1.8340 & 0.0195 & 1.8165 & 0.0020 & 1.8159 & 0.0014 \\
B7-B2 & 1.7873 & 1.8007 & 0.0134 & 1.7907 & 0.0034 & 1.7893 & 0.0020 \\
B5-B6 & 1.8028 & 1.8099 & 0.0071 & 1.8020 & -0.0008 & 1.8016 & -0.0012 \\
B3-B2 & 1.7589 & 1.7657 & 0.0068 & 1.7565 & -0.0024 & 1.7562 & -0.0027 \\
B4-B9 & 1.7231 & 1.7224 & -0.0007 & 1.7365 & 0.0134 & 1.7272 & 0.0041 \\
B4-B8 & 1.8016 & 1.7911& -0.0105 & 1.8146 & 0.0130 & 1.8064 & 0.0048 \\
B4-B1 & 1.7919 & 1.7968& 0.0049 & 1.8021 & 0.0102 & 1.7979 & 0.0060 \\
B4-B10 & 1.7805 & 1.7819 & 0.0014 & 1.7902 & 0.0097 & 1.7835 & 0.0030 \\
B4-B3 & 1.7781 & 1.7796& 0.0015 & 1.7848 & 0.0067 & 1.7815 & 0.0034 \\
B5-B10 & 1.9598 & 1.9585 & -0.0013 & 1.9649 & 0.0051 & 1.9640 & 0.0042 \\
B5-B2 & 1.8023& 1.7935& -0.0088 & 1.8026 & 0.0003 & 1.8028 & 0.0005 \\
B1-B2 & 1.7822 & 1.7735 & -0.0087 & 1.7813 & -0.0009 & 1.7815 & -0.0007 \\
B10$^\backprime$-B6 & 1.9604 & 1.9532 & -0.0072 & 1.9629 & 0.0025 & 1.9619 & 0.0015 \\
B8-B9 & 1.7936 & 1.7868 & -0.0068 & 1.7963 & 0.0027 & 1.7970 & 0.0034 \\
B5-B1 & 1.7573 & 1.7548 & -0.0025 & 1.7531 & -0.0042 & 1.7562 & -0.0011 \\
B3-B7 & 1.7569 & 1.7532 & -0.0037 & 1.7532 & -0.0037 & 1.7549 & -0.0020 \\
B8-B3 & 1.7503 & 1.7504 & 0.0001 & 1.7487 & -0.0016 & 1.7503 & 0.0000 \\
B4-(F/Br) &-& - & - & 1.3613&- & 1.9557& -\\
B7-F & - & 1.346020 & - & - &-&-&-\\
\bottomrule
\end{tabular}
\end{table}
Looking at Fig (\ref{bonds0fig}), we see that the shapes, for the ground state,  of the bond length changes are somehow similar for the substituted compounds at 4-F and 4-Br. However the latter bonds are mostly shorter than the former, which suggests that the substitution by Br leads to a more stable structure. As to the 7-F-substitution compound, it has the longest elongation, compared to the parent compound, exceeding 0.025 $A^o$.

\subsection{$HOMO$ and $LUMO$ Orbitals:}
The Highest Occupied Molecular Orbital (HOMO) and the Lowest Unoccupied Molecular Orbital (LUMO) 
play a crucial role in determining a molecule's photophysical properties, stability, and electronic characteristics. 
The energy gap (EG) between the HOMO and LUMO, commonly referred to as the band gap, is a key parameter 
that significantly affects the absorption and emission spectra, and is defined as: ($EG=E_{LUMO}-E_{HOMO}$).

The brominated molecule (4) exhibits the smallest energy gap, while the parent molecule (1) has the largest. 
For example, the fluorinated molecule at position B7 (2) shows an energy gap difference of approximately 0.03 eV compared to the fluorinated molecule 
at position B4. In contrast, the brominated molecule (4) has an energy gap that differs by about 0.42 eV from that of molecule (3). 
This suggests that the type of substituted atom has a significant effect on the energy gap, 
whereas the position of the substituted atom has a more subtle impact on \(EG\). 
These results highlight the critical role that both the type and position of substituted atoms play in determining molecular energy gaps. 
The energy gap (EG) values for the molecules studied are detailed in Table \ref{homolumo}, 
providing valuable insights into the electronic properties of these compounds.
\begin{table}[htbp]
\caption{HOMO, LUMO, and Energy Gap $(EG)$ at \ce{S0} geometry, with PBE0/def2-SVPD level of theory}
\label{homolumo}
\centering
\begin{tabular*}{\textwidth}{@{\extracolsep\fill}lccc}
\toprule
Compound               &  $HOMO$ $(ev)$   & $LUMO$ $(ev)$      & $EG$ $(ev)$  \\ \midrule
anti-\ce{B18H22}       & -7.963 [-7.874]& -3.203 [-3.104]& 4.760	[4.769]\\ 
7-F-anti-\ce{B18H21}   & -7.933 [-7.837]& -3.249 [-3.142]& 4.684	[4.696]\\ 
4-F-anti-\ce{B18H21}   & -7.983 [-7.882]& -3.327 [-3.208]& 4.656	[4.675]\\ 
4-Br-anti-\ce{B18H21}  & -7.650 [-7.671]& -3.407 [-3.282]& 4.243	[4.389]\\
\botrule
\end{tabular*}
\footnotetext{The values in brackets were calculated with CPCM(Hexane) model.}
\end{table}

\subsection{Vibrational Modes: IR and Raman Spectrum}
In order to determine the normal modes of vibration, one needs to compute the spectral eigenvalues of the free energy Hessian, which give the vibration frequencies. Minima correspond to all these frequencies being positive, whereas any negative value would correspond either to a maximum or a saddle point. In our case, we found all the vibrational frequencies are positive, whence energy stability is guaranteed. 

Frequency calculations were performed at the same theoretical level as that used for optimization. 
Each molecule exhibits 114 vibrational modes within the far and mid-infrared spectrum. 
These vibrational modes have been systematically categorized into four distinct spectral regions, 
each corresponding to specific molecular motions. 
The spectral range of 2500-3000 $cm^{-1}$ is associated with B-H bond stretching, 
while the B-H bond stretching within the B-H-B bridge is most prominent in the 2000-2500 $cm^{-1}$ region. 
Additionally, B-H bond bending within the B-H-B bridge is evident in the 1500-2000 $cm^{-1}$ interval. 
The region below 1500 $cm^{-1}$ encompasses more complex molecular motions and is recognized as the fingerprint region of the spectrum. 
Table \ref{modes} outlines the four spectral regions and the corresponding types of molecular motions.
\begin{table}[h]
\caption{Spectral Regions of Vibration Modes and Corresponding Molecular Motions}
\label{modes}
\centering
\begin{tabular*}{\textwidth}{@{\extracolsep\fill}lc}
\toprule
Frequency Range $(cm^{-1})$ & Type of motion \\ \midrule
2500-3000                   & B-H Bonds Stretching      \\      
2000-2500                   & B-H Bonds Stretching in B-H-B Bridge \\  
1500-2000                   & B-H Bonds Bending in B-H-B Bridge    \\ 
$<$1500                      & Complex Motions and Fingerprint Region \\
\botrule 
\end{tabular*}
\end{table}

The infrared (IR) spectrum of the halogenated molecules showed minimal deviations from that of the parent compound (1). This comes from the fact that the derivatives have the same 
general form of the configurations of laser borane anti-\ce{B18H22}, 
with the exception of an additional strong peak observed exclusively in the fluorinated derivatives (2) and (3). 
This peak is attributed to the stretching of the B-F bond. In contrast, the brominated derivative did not exhibit this peak. 
Comparative spectral analysis revealed that the IR spectrum of compound (4) is characterized by a more pronounced peak below 1500 $cm^{-1}$, 
a feature linked to B-Br bond stretching vibrations in this region, in addition to a stronger peak around 2200 $cm^{-1}$, due to B-H stretching in 
B-H-B bridges near the Br atom.  

\begin{figure}[H]
\caption{Calculated IR spectrum for all molecules}
\label{irall}
\centering
\includegraphics[width=\textwidth]{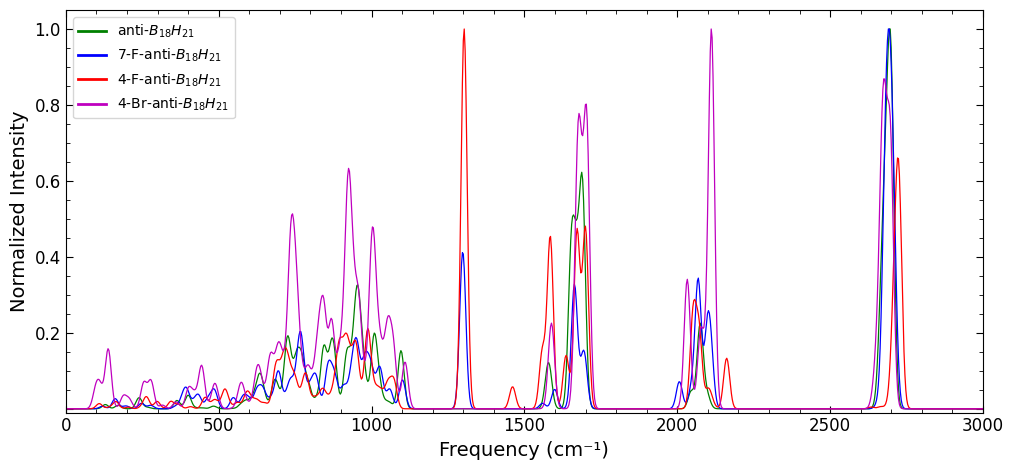}
\end{figure}

Figure \ref{irall} presents the IR spectra of all three derivatives alongside that of compound (1), providing a visual comparison 
that highlights both the similarities and differences in the spectra. 
The substitution of atoms within the molecule invariably affects the IR spectrum, 
as vibrational frequencies are sensitive to changes in atomic mass and bond strength. 
Consequently, changes in the type or position of substituted atoms can lead to significant shifts in these frequencies, 
thereby influencing the IR spectrum. As one can notice from figure \ref{irall}, compound (2) showed its strongest peak around 2700 $cm^{-1}$, 
which is assigned to B-H stretching, while compound (3) has the strongest peak around 1200 $cm^{-1}$, which is attributed to B-F elongation. This variation between position B7 for (2) and position B4 for (3) can be assigned to the higher electronegativity of B4, which leads to extra change in polarity, thus to a stronger peak in IR spectrum. We also show in Table (\ref{irtab}) the active vibrational IR modes for the four compounds, stating their frequencies and intensities confirming numerically what is noted qualitatively in figure (\ref{irall}).
\begin{table}[htbp]
\caption{Active Vibrational modes in IR spectra.}
\label{irtab}
\centering

\begin{tabular*}{\textwidth}{@{\extracolsep\fill}lcccccccc}
\toprule
\multirow{2}{*}{mode}& \multicolumn{2}{c}{anti-\ce{B18H22}}&\multicolumn{2}{c}{7-F-anti-\ce{B18H21}}
&\multicolumn{2}{c}{4-F-anti-\ce{B18H21}}&\multicolumn{2}{c}{4-Br-anti-\ce{B18H21}} \\
\cmidrule{2-9}
& \makecell{Freq \\ $(cm^{-1})$} & \makecell{Int \\ $(km/mol)$} & \makecell{Freq \\ $(cm^{-1})$} & \makecell{Int\\ $(km/mol)$} 
& \makecell{Freq\\ $(cm^{-1})$} & \makecell{Int\\ $(km/mol)$} & \makecell{Freq\\ $(cm^{-1})$} & \makecell{Int\\ $(km/mol)$} \\
\midrule

66 & 932.48 & 10.55 & 929.57 & 12.26 & 925.11 & 28.56 & 923.96 & 110.67 \\

92 & 1575.25 & 24.26 & 1298.5 & 305.49 & 1319.79 & 319.99 & 1110.64 & 13.39 \\

94 & 1653.6 & 144.87 & 1598.94 & 37.78 & 1585.87 & 47.17 & 1589.65 & 45.65 \\

95 & 1670.53 & 113.92 & 1658.71 & 20.53 & 1673.97 & 144.28 & 1673.61 & 137.36 \\

96 & 1689.17 & 165.41 & 1664.12 & 222.24 & 1679.75 & 44.74 & 1684.9 & 88.09 \\

97 & 1693.1 & 42.46 & 1690.33 & 58.66 & 1698.08 & 148.39 & 1703.79 & 188.02 \\

98 & 2045.41 & 15.94 & 1698.38 & 62.15 & 1706.76 & 150.49 & 1714.71 & 1.66 \\

100 & 2074.89 & 79.98 & 2046.78 & 37.77 & 2053.28 & 75.59 & 2035.36 & 29.6 \\

101 & 2078.68 & 0.24 & 2068.52 & 249.14 & 2073.83 & 62.14 & 2075.93 & 48.06 \\

104 & 2660.43 & 11.79 & 2110.38 & 88.27 & 2104.39 & 42.61 & 2112.13 & 253.82 \\

116 & 2694.28 & 42.07 & 2695.76 & 251.49 & 2689.84 & 45.88 & 2691.25 & 25.26 \\

119 & 2702.98 & 100.37 & 2710.57 & 169.12 & 2702.35 & 34.74 & 2699.53 & 68.59 \\
\bottomrule
\end{tabular*}
\end{table}

\begin{figure}[H]
\caption{Calculated Raman Activity Spectrum for all four Molecules}
\label{ramanact}
\centering
\includegraphics[width=\textwidth]{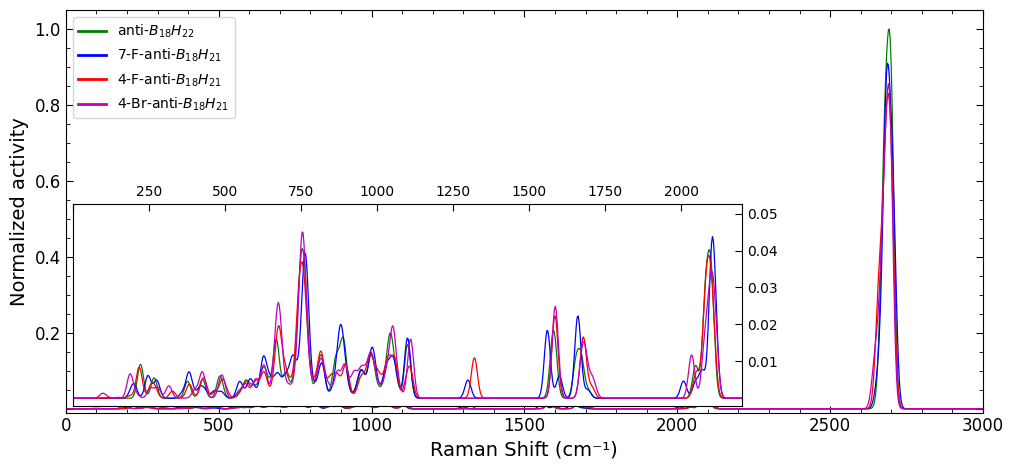}
\end{figure}

Numerical frequency calculations were performed, leading to the determination of Raman activities. 
Raman activity is the difference in intensity between stoke's (Photon loss energy) and anti-stoke's (Photon gain energy) scattering.
All molecules exhibited a prominent peak within the 2600-2700 $cm^{-1}$ spectral range, corresponding to various symmetrical stretching of the B-H bonds, 
However, all other modes were not active in Raman spectra, a fact which arises because most of these modes are bending or asymmetrical stretching,  
leading thus to an insufficient change in molecule polarizability.  
Actually, in the infrared spectrum, a distinct peak was observed for fluorinated molecules, attributed to B-F stretching vibrations, 
though it displayed minimal Raman activity. Notably, this peak is absent in both the parent molecule (1) and the brominated molecule (4). 
Figure \ref{ramanact} presents the complete Raman Activity spectrum in comparison with molecule (1). 

\begin{figure}[H]
\caption{Calculated Raman Intensity Spectrum for All Molecules, Laser Radiation= 18797 $cm^{-1}$ at 298.15 $K$}
\label{ramanabs}
\centering
\includegraphics[width=\textwidth]{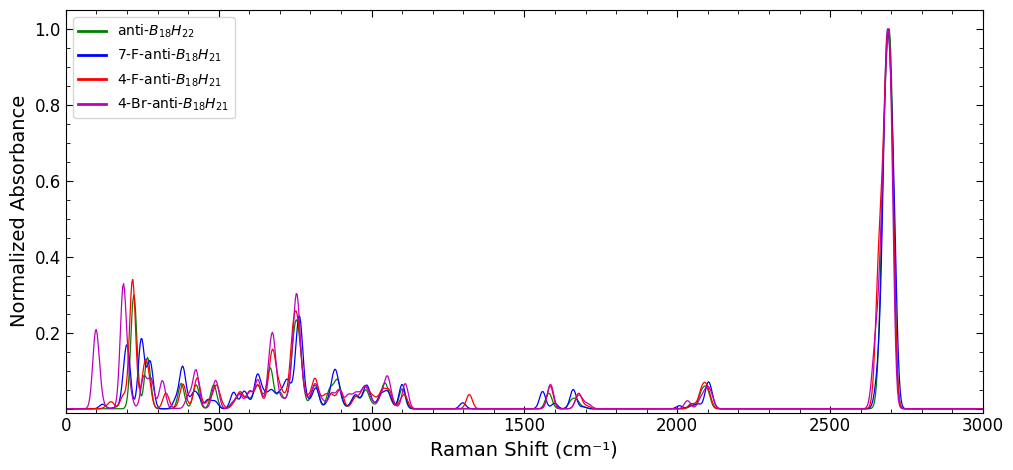}
\end{figure}

The Raman intensity data were calculated using Chemcraft software, with laser radiation at 18797 $cm^{-1}$, 
which is the optimal frequency for investigating inorganic molecules, at a temperature of 298.15 K.
Upon comparing the Raman activity spectrum, reliant on the Raman scattering intensity difference between left- and right-circularly polarized light due to molecular chirality, several peaks exhibited enhanced intensity in the Raman intensity spectrum, particularly 
in the region below 1200 $cm^{-1}$. These changes indicate increased favourability of the corresponding vibrational modes at the selected laser wavelength. 
Moreover, looking at the relation between activity $S_i$ and intensity $R_i$ in Raman spectra (Eq. \ref{ram}, where $\nu_i$ is the frequency of the $i_{th}$ mode, $\nu_0$ is the laser frequency,  and $K, T, c, h$ are Boltzman's constant, 
temperature, speed of light and Planck's constant, respectively)
\begin{equation}
R_i= \frac{(2 \pi)^2}{45} (\nu_0-\nu_i)^4 \times \frac{h}{8 \pi^2 c \nu_i(1-e^{-\frac{h \nu_i c}{KT}})} \times S_i
\label{ram}
\end{equation}
 justifies the existence of intense peaks below 1500 $cm^{-1}$, despite their weak 
activity, since $\nu$ being in the denominator leads to large Raman intensity at smaller frequencies regardless of the activity. Furthermore, compound (4) showed a unique peak around 100 $cm^{-1}$, which can be attributed to the movement of the heavy bromine atom. 
Figure \ref{ramanabs} displays the Raman intensity spectrum for the studied molecules. 

Finally, the frequency calculations provided thermodynamic properties (covering the electronic energy, entropy, enthalpy, Gibbs, and vibrational heat capacity), which are summarized in Table \ref{thermotab}.

Thermodynamic properties play an important role in determining the thermal stability of compounds, where those of negative enthalpy tend to be more stable, which would make them useful in optoelectronic devices working at high temperatures. Table \ref{thermotab} shows a clear increase in Enthalpy and in Gibbs free energy in halogenated compounds, in particular those corresponding to Bromine  substitution, predicting thus a better thermal stability.  

The slight differences in entropy values are due to the original Borane structure being kept by the substituted compounds, which lead to similar vibrational entropies. The largest vibrational entropy  leads to coupling between the electronic and vibrational levels, which enhances the non-radiative relaxation processes competing with the radiative ones. As to the large differences in Enthalpy and Gibbs free energy, they are caused by the large discrepancies in the compounds electronic energies, which result from the increasing interaction of electrons with nuclei and other electrons in the presence of halogens, the most pronounced of which corresponds to Bromine substitution.  The higher electronic energy in the bromine-substituted compound is due to its
heavy bromine atom, including a larger number of electrons leading, thus, to more interactions amidst electrons and nuclei. A larger absolute value of electronic energy means a more challenging task of breaking up the compound, thus a better stability, although other factors, such as bond lengths and formation energy, play also a role. 

\begin{table}[h]
\centering
\caption{Thermodynamics Properties at $T$ = 298.15 $K$ and $P$ = 1 $atm$ }\label{thermotab}
\begin{tabular*}{\textwidth}{@{\extracolsep\fill}lcccccc}
\toprule
Molecule &\makecell{Electronic \\$E_{el}$ $(Eh)$}&\makecell{Entropy\\$(Eh)$}&\makecell{Enthalpy\\$(Eh)$}&\makecell{Gibbs\\ $G$ $(Eh)$ }
&\makecell{ $G-E_{el}$ \\$(Eh)$} &\makecell{ $C_{V,vib}$ \\ $(J/K.mole)$}\\
\midrule
anti-\ce{B18H22} & -459.78178 & 0.05091 & -459.469371 & -459.52028& 0.26150        & 280.24  \\ 
7-F-anti-\ce{B18H21} & -558.90979 & 0.05306 & -558.60264& -558.65571 & 0.25408     & 291.55  \\ 
4-F-anti-\ce{B18H21} & -558.90578 & 0.05295 & -558.59895& -558.65190& 0.25387      & 292.52  \\ 
4-Br-anti-\ce{B18H21} & -3032.76923 & 0.05526 & -3032.46356& -3032.51883 & 0.25039 & 296.96  \\
\bottomrule
\end{tabular*}
\end{table}

The vibrational heat capacity was computed using the following statistical formula:
\begin{eqnarray}
C_{V,vib}&=&R \sum_{i=1}^{\alpha}\left[\left(\frac{h \nu_i}{K_B T}\right)^2 \frac{e^{-\frac{h \nu_i}{K_B T}}}{\left(1-e^{-\frac{h \nu_i}{K_B T}}\right)^2}\right]
\end{eqnarray}
where, $\alpha$ is the number of vibrational modes, $\nu_i$ is the frequency of the $i^{th}$ mode, T is the temperature and 
$R$, $K_B$, and $h$ are ideal gas, Boltzman, and Planck constants, respectively.
Heat capacities play an essential role in optoelectronic devices, where
materials with higher values can manage thermal changes better,
maintaining stable performance under varying operating conditions.

\subsection{TD-DFT Calculation:}
In computational chemistry, exploring excited states is crucial for understanding the photophysical properties of compounds. 
Time-dependent density functional theory (TDDFT) is an essential tool for simulating electronic transitions. 
In this study, we employed TDDFT to evaluate the optimized geometry of the first singlet excited state (\(S_1\)) 
and to calculate the vertical absorption and emission energies, offering theoretical insights that correspond well with experimental findings. 
We utilized two methods, PBE0/def2-SVPD and B3LYP/6-311+G(d), in both the gas phase and solvent phase using Hexane with the CPCM model.
\subsubsection{Vertical Absorption Energies}
In our research, we conducted TD-DFT calculations to validate the vertical absorption energies of excited states. 
The results for the parent compound (1) closely matched experimental data on absorption wavelengths, 
although a discrepancy was observed in the oscillator strength compared to previous calculations. 
The oscillator strength for the transition to the second excited state (\(S_2\)) was minimal. 
However, the oscillator strength for the initial absorption aligned well with values calculated using the more accurate CASPT2 method. 
Incorporating solvent effects with the CPCM (Hexane) model increased the probability of the transition, though the transition to \(S_2\) remained negligible. 
Table \ref{antiuvtab} presents the first three transitions, comparing them to results from previous research \cite{anti2012}.
\begin{table}[htpb]
\caption{Vertical absorption energy for anti-\ce{B18H22} in comparison with previous work \cite{anti2012}}
\label{antiuvtab}
\centering
\begin{tabular*}{\textwidth}{@{\extracolsep\fill}lccccccc}
\toprule
\multirow{2}{*}{State} &\multicolumn{2}{c}{\makecell{PBE0\\def2-SVPD \footnotemark[1]}} & 
\multicolumn{2}{c}{\makecell{B3LYP \\6-311+G(d) \footnotemark[1]}} & 
\multicolumn{2}{c}{\makecell{CASPT2 \cite{anti2012} \footnotemark[2]}} & \multicolumn{1}{c}{\makecell{Exp \cite{anti2012}}}  \\
\cmidrule(lr){2-3} \cmidrule(lr){4-5} \cmidrule(lr){6-7} \cmidrule(lr){8-8}
& $\lambda_{VA} (nm)$ & $f$& $\lambda_{VA} (nm)$ & $f$ & $\lambda_{VA} (nm)$ & $f$ & $\lambda_{abs} (nm)$\\
\midrule
$S_1$ & 308 [310]  & 0.143 [0.201]      &311 [313]   &0.141 [0.196]  & 315 & 0.265&329 \\
$S_2$ & 263 [262]  & 0.000 [0.000]      &264 [263]   &0.000 [0.000]  & 248 & 0.051&272 \\
$S_3$ & 259 [259]  & 0.055 [0.084]      &261 [261]   &0.056 [0.084]  & 212 & 0.884&215 \\
\bottomrule
\end{tabular*}
\footnotetext{$\lambda_{VA}$ is the calculated vertical absorption wavelength at $S_0$ geometry, $f$ is the oscillator strength. 
$\lambda_{abs}$ is the experimental absorption wavelength observed in hexane solution. 
The values in brackets were calculated with CPCM model in Hexane}
\footnotetext[1]{As obtained in ORCA 5.0 (this work).}
\footnotetext[2]{As obtained in MOLCAS 7.0 (previous work \cite{anti2012}).} 
\end{table}
Also, we show in figure (\ref{uvanti}) the calculated UV-vis spectra for the anti-\ce{B18H22} compound compared to previous work  \cite{anti2012} quoting both calculations done with CASPT2 in MOLCAS software and measurements in hexane.
 
\begin{table}[htbp]
\centering
\captionof{figure}{Calculated UV-vis spectra for anti-\ce{B18H22} in comparison with previous work}
\label{uvanti}
\begin{tabular}{c}
\includegraphics[width=\textwidth]{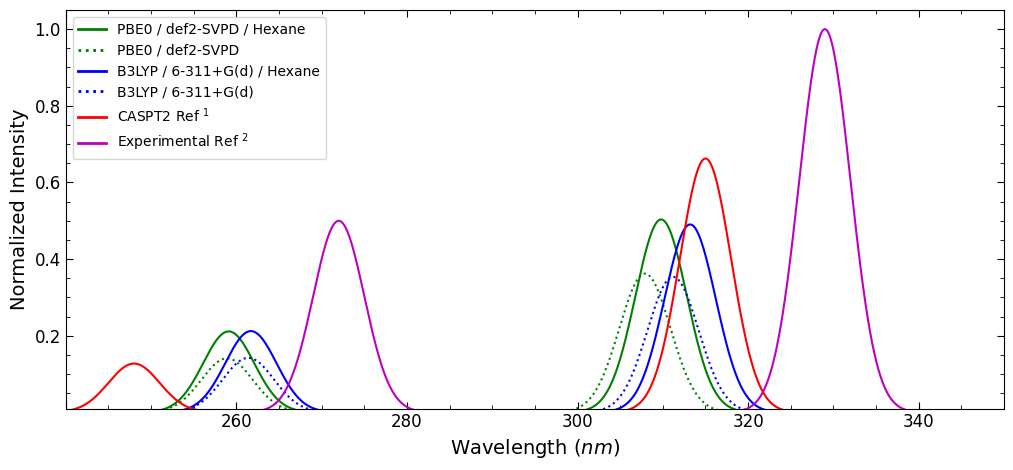}
\end{tabular}
\footnotetext[1]{Calculated at CASPT2 level of theory on MOLCAS software \cite{anti2012}.}
\footnotetext[2]{Measure in hexane \cite{anti2012}.}
\end{table}

As to the fluorinated molecules, their transition energies showed minimal deviation from those of compound (1), with a slight redshift of approximately 8 nm. 
The oscillator strength values for compound (2) were similar to those of compound (1). 
In contrast, compound (3) exhibited a significantly increased oscillator strength for the transition to the \(S_2\) state, 
which was negligible in both compounds (1) and (2), while compound (2) (Flourinated at B7) showed the highest $f$ value among all studied molecules. This indicates an enhancement 
in molar absorptivity, which is compatible with the highest absorbency when substituted with Cl at position B7 in a previous study \cite{cl2022}. 
Theoretically, it is worth mentioning that the larger the oscillator strength {\it f} for specific transition between two states is, 
the more similar  the corresponding wave functions are. 

On the other hand, using CPCM (Hexane) model enhanced the transition probabilities, as reflected in the oscillator strength values shown in Table \ref{74fuv}, 
which compares the first three vertical absorption wavelengths for the fluorinated molecules (2) and (3) with those of the parent molecule (1). 

Actually, the transition probability of the main absorption peak, corresponding to vertical transition from S0 to S1, increased in Hexane of about 40\% for compounds (1), (2), and (3), whereas significant enhancement occured for the compound (4) reaching about 90\% due to more polarity caused by bromine atom. This suggests a solvent-cluster interaction of the Van der Waals-type, since the solvent is non-polar.   

\begin{table}[htbp]
\caption{Vertical absorption energy for flourinated molecules compared to parent compound}
\label{74fuv}
\centering

\begin{tabular*}{\textwidth}{@{\extracolsep\fill}lcccccc}
\toprule
\multirow{2}{*}{State} & \multicolumn{2}{c}{anti-\ce{B18H22}} & \multicolumn{2}{c}{7-F-anti-\ce{B18H21}} & \multicolumn{2}{c}{4-F-anti-\ce{B18H21}} \\ 
 \cmidrule(lr){2-3} \cmidrule(lr){4-5} \cmidrule{6-7}
&$\lambda_{VA} (nm)$&$f$&$\lambda_{VA} (nm)$&$f$&$\lambda_{VA} (nm)$& $f$ \\
\midrule
&&&\makecell{PBE0\\def2-SVPD}&  &&\\
\midrule
$S_1$ & 308 [310]& 0.1448 [0.2014]& 312 [315]& 0.1637 [0.2229]& 316 [317]& 0.1457 [0.2022]\\ 
$S_2$ & 263 [262]& 0.0000 [0.0000]& 262 [262]& 0.0009 [0.0014]& 272 [271]& 0.0533 [0.0774]\\ 
$S_3$ & 259 [259]& 0.0564 [0.0846]& 256 [256]& 0.0470 [0.0720]& 261 [261]& 0.0176 [0.0240]\\
\midrule		  
& &&\makecell{B3LYP\\6-311+G(d)}& &&\\
\midrule
$S_1$ & 311 [313]& 0.1415 [0.1962]& 316 [318]& 0.1637 [0.2157]& 320 [321]& 0.1407 [0.1951]\\ 
$S_2$ & 264 [263]& 0.0000 [0.0000]& 264 [263]& 0.0006 [0.0011]& 275 [274]& 0.0539 [0.0778]\\ 
$S_3$ & 261 [261]& 0.0569 [0.0849]& 259 [259]& 0.0468 [0.0715]& 264 [264]& 0.0138 [0.0189]\\
\bottomrule
\end{tabular*}
\footnotetext{$\lambda_{VA}$ is the calculated vertical absorption wavelength at $S_0$ geometry, $f$ is the oscillator strength. 
No previous experimental nor theoretical data available for those two new derivatives. Again, the values in brackets were calculated 
with CPCM model in Hexane solvent.}
\end{table}

Likewise, we show in figure (\ref{uvanti7f4fref}) our calculated UV-vis spectra for the fluorinated derivatives compared to halogenated compounds at B7 position previously quoted (\cite{i2019} for I and \cite{cl2022} for Cl).

\begin{table}[htbp]
\centering
\captionof{figure}{Calculated UV-vis spectra for flourinated derivatives in comparison with 
the previous halgenated at position B7}
\label{uvanti7f4fref}
\begin{tabular}{c}
\includegraphics[width=\textwidth]{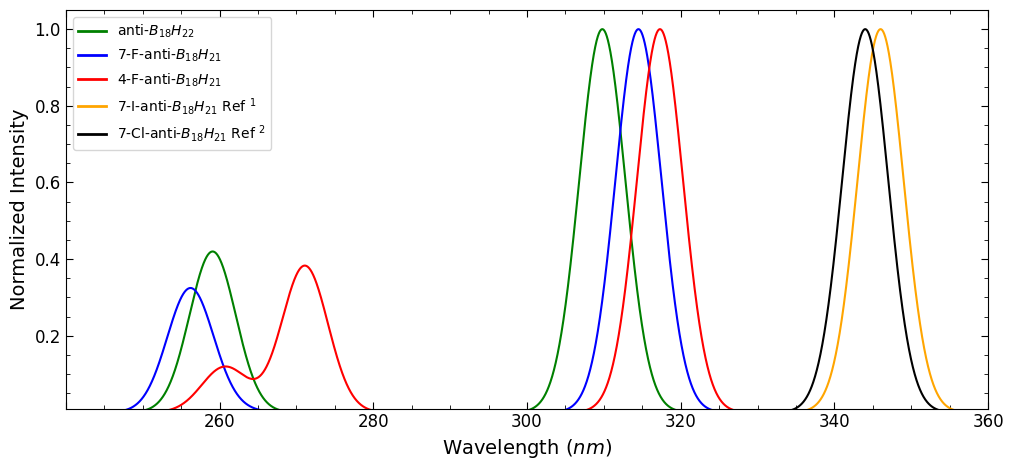}
\end{tabular}
\footnotetext{calculated at PBE0/def2-SVPD/Hexane level of theory. Normalized for the highest oscillator strength for each spectra, 
separately}
\footnotetext[1]{Previous halogenated 7-I-anti-\ce{B18H21} \cite{i2019}.}
\footnotetext[2]{Previous halogenated 7-Cl-anti-\ce{B18H21} \cite{cl2022}.}
\end{table}

The brominated molecule exhibited a redshift in its absorption wavelength of approximately 47 nm, 
accompanied by a lower oscillator strength compared to the parent compound (1). The bromine-substituted compound exhibits a greater redshift than the
fluorine-substituted ones, which signifies a reduction in
the energy required to excite the former molecule.
The calculated vertical absorption energies closely matched the experimental data \cite{br2020}. 
Using the PBE0/def2-SVPD method, the first transition to \(S_1\) at 354 nm corresponds well with the experimental peak observed at 341 nm, 
while the transition to \(S_2\) at 295 nm aligns with the peak around 300 nm. 
When using the B3LYP/6-311+G(d) method, the transitions were redshifted but still in good agreement with experimental observations. 
The best agreement was achieved by using the CPCM (Hexane) model with the PBE0/def2-SVPD method. 
Table \ref{4bruvtab} presents the vertical absorption wavelengths, comparing them with previous experimental and calculated data.
As illustrated in Figure \ref{4bruvtab}, our calculated values align more closely with the experimental data, 
even though we did not include relativistic effects, which were considered in previous studies \cite{br2020}.
\begin{table}[htbp]
\caption{Vertical absorption energies for 4-Br-anti-\ce{B18H21}}
\label{4bruvtab}
\centering
\begin{tabular*}{\textwidth}{@{\extracolsep\fill}lccccccc}
\toprule
\multirow{2}{*}{State} &\multicolumn{2}{c}{\makecell{PBE0\\def2-SVPD \footnotemark[1]}} & 
\multicolumn{2}{c}{\makecell{B3LYP\\6-311+G(d)\footnotemark[1]}} 
& \multicolumn{2}{c}{\makecell{PBE0\\DZP \cite{br2020}\footnotemark[2]}} & \multicolumn{1}{c}{Exp \cite{br2020}}  \\
\cmidrule(lr){2-3} \cmidrule(lr){4-5} \cmidrule(lr){6-7} \cmidrule(lr){7-7}
 & $\lambda_{VA} (nm)$ & $f$ & $\lambda_{VA} (nm)$ & $f$ &$\lambda_{VA} (nm)$ & $f$ & $\lambda_{abs} (nm)$\\
\midrule
$S_1$ & 354 [341]& 0.0627  [0.1189]&363 [349]&  0.0527 [0.1019]& 357 & 0.0443 &343 \\
$S_2$ & 341 [325]& 0.0367  [0.0514]&351 [334]&  0.0341 [0.0472]& 347 & 0.0336 &343 \\
$S_3$ & 295 [291]& 0.0809  [0.0829]&299 [296]&  0.0861 [0.0928]& 288 & 0.0909 &300 \\
\bottomrule
\end{tabular*}
\footnotetext{$\lambda_{VA}$ is the calculated vertical absorption wavelength at $S_0$ geometry, $f$ is the oscillator strength. 
$\lambda_{abs}$ is the experimental absorption wavelength in cyclohexane solution. 
The values in brackets were calculated with CPCM model in Hexane.}
\footnotetext[1]{As obtained  in ORCA software without Relativistic effects (this work).}
\footnotetext[2]{As obtained in ADF with Relativistic effects (previous worke \cite{br2020}).}
\end{table} 

\begin{table}[htbp]
\centering
\captionof{figure}{Calculated UV-vis for brominated derivative in comparison with previous work \cite{br2020}.}
\label{4bruv}
\begin{tabular}{c}
\includegraphics[width=\textwidth]{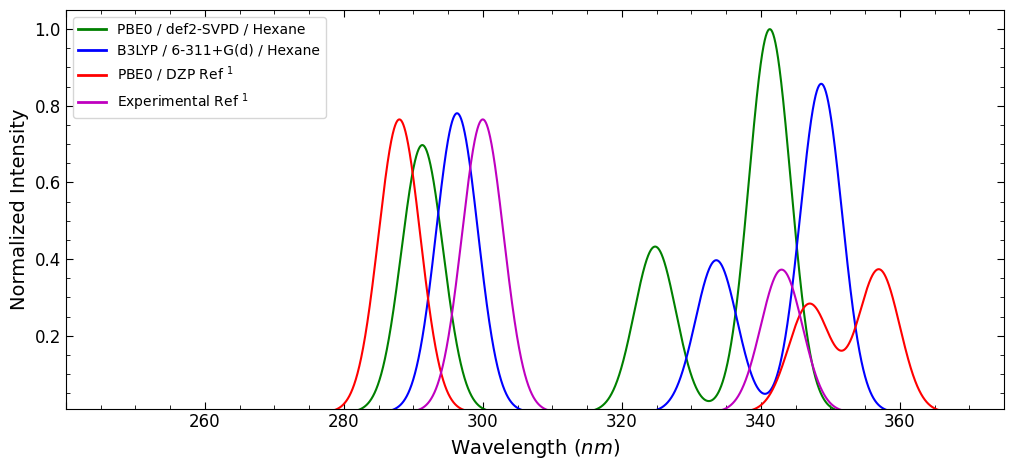}
\end{tabular}
\footnotetext[1]{Calculated with scalar relativistic effects in ADF software. Measured in cyclohexane \cite{br2020}.}
\end{table}

\subsubsection{Vertical Emission Energies}
We optimized the geometry of the first singlet excited state (\ce{S1}) for each molecule. For compounds (1) and (2), 
the most significant elongation was observed in the B9-B10 and B9$^\backprime$-B10$^\backprime$ bonds, with a change of approximately 0.05 Å. 
The most substantial contraction occurred in the B6-B10$^\backprime$ bond, with a difference of around 0.1 Å. 
In compounds (3) and (4), the B9-B10 bond exhibited the maximum elongation, with a change of approximately 0.09 Å. 
Compound (3) also showed maximum contraction in the B6-B10$^\backprime$ bond with a difference of about 0.1 Å. 
For compound (4), the B5-B10 and B7-B8 bonds experienced the most significant contractions, with changes of approximately 0.12 Å. 
Table \ref{bonds1} provides a detailed comparison of the top three changes in B-B bond lengths between the optimized geometries of \ce{S1} and \ce{S0}.

\begin{figure}
\caption{Maximal differences $\delta$ in B-B bonds between halogenated derivatives and parent compound, at their $S_1$ geometries}
\label{bonds1fig}
\centering
\includegraphics[width=\linewidth]{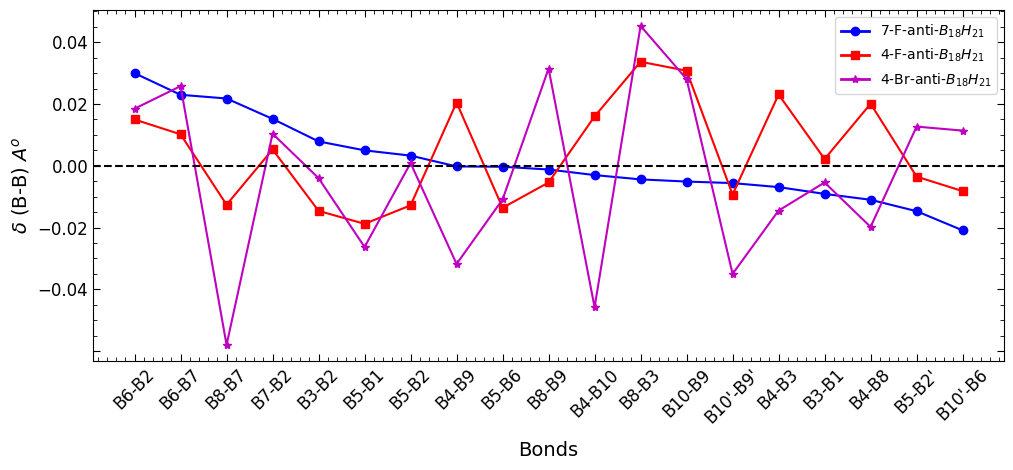}
\end{figure}

\begin{figure}
\centering
\includegraphics[width=\linewidth]{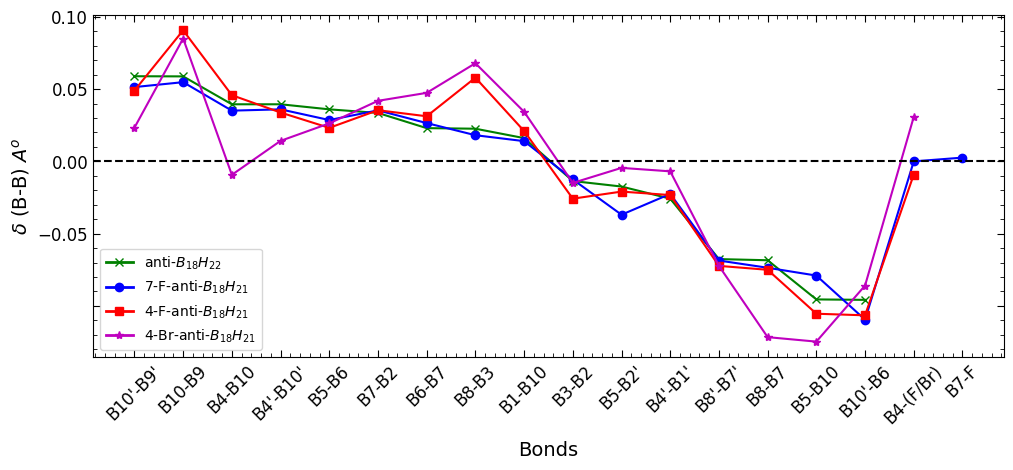}
\caption{Maximal differences $\delta$ in B-B bonds between $S_1$ and $S_0$ geometries for each molecule under this study.}
\label{bonds1s0}
\end{figure}

\begin{table}[h]
\caption{Differences, $\delta$ in B-B bonds length between \ce{S1} and \ce{S0} geometries}
\label{bonds1}
\centering
\begin{tabular*}{\textwidth}{@{\extracolsep\fill}lcccc}
\toprule
\multirow{2}{*}{Bond} & anti-\ce{B18H22} & 7-Fanti-\ce{B18H21}& 4-F-\ch{anti-B18H21} & 4-Br-anti-\ce{B18H21} \\
\cmidrule{2-5}
& $\delta(\ce{s1-s0}) \, A^o$ & $\delta(\ce{s1-s0}) \, A^o$ & $\delta(\ce{s1-s0}) \, A^o$ & $\delta(\ce{s1-s0}) \, A^o$ \\
\cmidrule{1-5}
B5-B10 & -0.0955 & -0.079 & -0.1054 & -0.1247 \\
B7-B8 & -0.0684 & -0.0736 & -0.075 & -0.1216 \\
B6-B10$^\backprime$ & -0.0958 & -0.1095 & -0.1065 & -0.0859 \\
B4-B10 & 0.0395 & 0.0351 & 0.0458 & -0.0093 \\
B4$^\backprime$-B10$^\backprime$ & 0.0395 & 0.036 & 0.0338 & 0.0142 \\
B9$^\backprime$-B10$^\backprime$ & 0.0589 & 0.0514 & 0.049 & 0.0233 \\
B7-B6 & 0.023 & 0.0265 & 0.0312 & 0.0474 \\
B3-B8 & 0.0226 & 0.0181 & 0.0579 & 0.0679 \\
B9-B10 & 0.0588 & 0.0548 & 0.0907 & 0.085 \\
B7-F   &     -   &0.0026 &   -& -\\
B4-F &-&-&-0.0091&-\\
B4-Br&-&-&-&0.0308\\
\bottomrule
\end{tabular*}
\end{table}
In figure (\ref{bonds1fig}), we show the maximal deviations in B-B bonds between halogenated derivatives and the parent compound. Likewise, figure (\ref{bonds1s0}) shows these deviations between the $S_1$ and $S_0$ geometries for each studied molecule. 

Additionally, the energies of the HOMO and LUMO orbitals changed at the \ce{S1} optimized geometry, resulting in a general decrease in the energy gap (EG). 
Compounds (1) and (3) exhibited an EG difference of approximately -0.8 eV, while compounds (3) and (4) showed a change in EG values of approximately -0.9 eV. 
These changes contribute to a redshift in emission energy, as evidenced by the vertical emission energy calculations. 
The EG values for all studied molecules at their \ce{S0} and \ce{S1} geometries are presented in Table \ref{homolumos0s1}.

Energy gaps play an important role in determining the compounds optical physical properties, and one can consider them as first approximations of excitation energies. The smaller the energy gap is, the larger the redshift in absorption wavelength becomes, a fact which would contribute to diminishing the energy required to excite the compounds.  

\begin{table}[h]
\caption{HOMO, LUMO and EG energy gap at \ce{S1} geometry, with PBE0/def2-SVPD level of theory.}
\label{homolumos0s1}
\centering
\begin{tabular*}{\textwidth}{@{\extracolsep\fill}lccc}
\toprule
Compound               &  $HOMO$ $(ev)$   & $LUMO$ $(ev)$      & $EG$ $(ev)$  \\ 
\midrule
anti-\ce{B18H22}     & -7.674 [-7.585]& -3.760 [-3.661]&3.914 [3.925]\\ 
7-F-anti-\ce{B18H21} & -7.645 [-7.550]& -3.815 [-3.707]&3.830 [3.843]\\ 
4-F-anti-\ce{B18H21} & -7.657 [-7.562]& -3.914 [-3.795]&3.743 [3.767]\\ 
4-Br-anti-\ce{B18H21}& -7.460 [-7.520]& -4.097 [-3.969]&3.363 [3.551]\\
\bottomrule
\end{tabular*}
\footnotetext{The values in brackets were calculated with CPCM(Hexane) model.}
\end{table}

Based on the optimized \ce{S1} geometry, we conducted time-dependent density functional theory (TD-DFT) calculations to determine 
the vertical emission energies. Figure \ref{absemall} displays the normalized absorption and emission spectra for all molecules calculated at PBE0/def2-SVPD/Hexane level of theory, 
in comparison with previous data (\cite{anti2012} for hexane solvent and \cite{br2020} for cyclohexane).
\begin{table}[htbp]
\centering
\captionof{figure}{Absorption and Emission spectra calculated at PBE0/def2-SVPD/Hexane level of theory, 
in comparison with previous data}
\label{absemall}
\begin{tabular}{c}
\includegraphics[width=\textwidth]{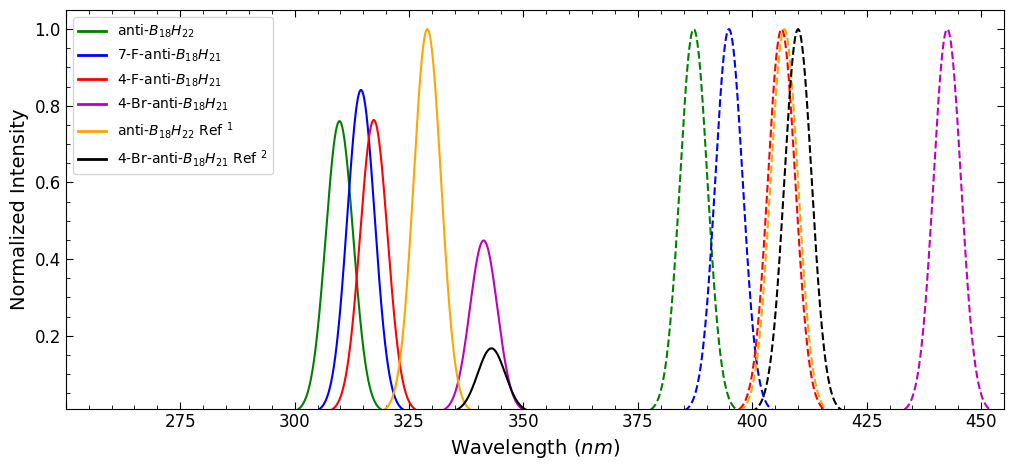}
\end{tabular}
\footnotetext{Absorption denoted as solid line, emission as dashed line, and previous work in orange, 
and black for compound (1), and (4), respectively. 
Absorption intensities were normalized for the max oscillator strength of compound (1) in the previous work \cite{anti2012}, 
whereas emission intensities were normalized separately.}
\footnotetext[1]{In hexane \cite{anti2012}.}
\footnotetext[2]{In cyclohexane \cite{br2020}.}
\end{table}
 
\begin{table}[htbp]
\caption{Vertical emission wavelength and radiative lifetime.}
\label{emtau}
\begin{tabular*}{\textwidth}{@{\extracolsep\fill}lcccccc}
\toprule
Compound & $\lambda_{VA}$ $(nm)$ & $\lambda_{VE}$ $(nm)$ & $\tau$ ($ns$) &$\lambda_{abs}$ $(nm)$ & $\lambda_{F}$ $(nm)$& $\tau_F$ ($ns$) \\
\midrule
&&\makecell{PBE0\\def2-SVPD}&& &Exp &\\
\midrule
anti-\ce{B18H22}     & 308 [310]& 383 [387]& 13.7 [10.5]& 329 \footnotemark[1]&407 \footnotemark[1]&11.2\footnotemark[1] \\ 
7-F-anti-\ce{B18H21} & 312 [315]& 390 [394]& 13.4 [10.4]& -  & -& -     \\ 
4-F-anti-\ce{B18H21} & 316 [317]& 404 [406]& 14.8 [11.3]& -  & -& -     \\
4-Br-anti-\ce{B18H21}& 354 [341]& 473 [442]& 78.5 [34.4]& 341\footnotemark[2]&410\footnotemark[2] &10.6\footnotemark[2] \\
\midrule		  
& &\makecell{B3LYP\\6-311+G(d)}&& &Theo &\\
\midrule
anti-\ce{B18H22}      & 311 [313]& 386 [390]& 13.91 [10.7]& 315 \footnotemark[1]&426\footnotemark[1] &5.6\footnotemark[1] \\ 
7-F-anti-\ce{B18H21}  & 316 [318]& 394 [399]& 13.75 [10.7]& -  & -& -     \\ 
4-F-anti-\ce{B18H21}  & 320 [321]& 408 [410]& 15.38 [11.7]& -  & -& -     \\
4-Br-anti-\ce{B18H21} & 363 [349]& 490 [457]& 101.9 [44.1]& 357\footnotemark[2]&374\footnotemark[2] &- \\
\bottomrule
\end{tabular*}
\footnotetext{$\lambda_{VA}$ vertical absorption wavelength, $\lambda_{VE}$ vertical emission wavelength, $\tau$ radiative lifetime in nanosecond. $\lambda_{abs}$ absorption wavelength, $\lambda_F$ flourescence wavelngth and $\tau_F$ flourescence lifetime. Exp experimental and Theo theoretical previous values. The values in bracket were calculated with CPCM(Hexane) model.}
\footnotetext[1]{Experimental values measured in hexane, and calculated at CASPT2 level of theory \cite{anti2012}.}
\footnotetext[2]{Measured in cyclohexane, and calculated at PBE0/DZP with scalar relativistic effect \cite{br2020}}
\end{table}

 Our analysis revealed that all compounds emitted light within the visible spectrum. 
Notably, the fluorinated derivatives (2) and (3) exhibited slight redshifts, while the brominated derivative (4) demonstrated a significant redshift 
of approximately 100 nm compared to compound (1).

Compound (1) has a vertical emission wavelength (\(\lambda_{VE}\)) of 383 nm, showing a redshift of 75 nm 
from the vertical absorption wavelength (\(\lambda_{VA}\)). 
This result aligns well with the observed peak around 407 nm and the calculated peak of 426 nm at the CASPT2 level of theory \cite{anti2012}. 
Furthermore, the fluorinated molecules (2) and (3) displayed vertical emissions at 390 nm and 404 nm, respectively, 
with redshifts of 78 nm and 87 nm. The brominated molecule exhibited the highest vertical emission wavelength at 473 nm, 
although this value deviates significantly from both experimental and calculated data.
 
We hypothesize that this discrepancy may result from the exclusion of relativistic effects associated with the heavy bromine atom. 
Including a solvent model (CPCM in Hexane) improved the agreement with previous experimental and theoretical values, 
especially for the brominated derivative (4). For radiative lifetime, calculated using the Strickler-Berg equation \cite{strickler1962}, 
the values were in good agreement for compound (1) with the experimental data, but an overestimation was observed for the brominated compound (4). 
However, closer values were obtained by incorporating the CPCM (Hexane) model, as shown in Table \ref{emtau}, 
which details the vertical emission energies and the excited state's lifetime. It is worth mentioning that compound (1), which has laser emission, 
is not the one with the highest $\tau$ values. This comes since longer radiative lifetimes might lead to other processes occurring, which can negatively affect the efficiency 
of the laser. 

The overestimation of the excited state lifetime using the ``Strickler-Berg" formula:
\begin{eqnarray}
\tau &=&\frac{1}{2.142005 E_{\mbox{\tiny VE}}^3 {\mbox{TDM}^2}}
\end{eqnarray}
is due to two factors. First, the transition dipole moment $\mbox{TDM}$, related to oscillator strength, is of small value in Bromine substituted compound, which, due to its existence in the denominator, was argued to give overestimating values \cite{strickler1962}. Second, the vertical emission energy $E_{\mbox{\tiny VE}}$ enters the formula raised to the cubic power making it of paramount effect. The fact that the estimation was done for emission wavelength at 442 nm, whereas the experimental one is at 410 nm, leads to reducing  $E_{\mbox{\tiny VE}}$ and overestimating  the lifetime. Moreover, neglecting relativistic effects led to an additional underestimation (overestimation) of $E_{\mbox{\tiny VE}}$ ($\tau$).

\subsubsection{Adiabatic and 0-0 Energies}
Table \ref{adia} shows the adiabatic energies $\lambda_{adia}$ and origin band energies $\lambda_{0-0}$. 
Adiabatic energy is the difference between \ce{S0} and \ce{S1} energies at their respective geometries, 
while origin  band energy is the energy difference between the lowest vibrational energy state in $S_0$ and that for $S_1$, 
which can be calculated by adding the difference in zero-point vibration energy between \ce{S0} and \ce{S1}, to the adiabatic energy \cite{e00}. 
The origin band energy calculated in this study is in good agreement with the experimental value 364 $nm$ and better than the theoretical value calculated 
at CASPT2 level of theory 341 $nm$ \cite{anti2012}.

\begin{table}[htbp]
\caption{Adiabatic and origin band energies.}
\label{adia}
\begin{tabular*}{\textwidth}{@{\extracolsep\fill}lccccc}
\toprule
\multirow{2}{*}{Compund} & \multicolumn{2}{c}{PBE0 / def2-SVPD} & \multicolumn{2}{c}{B3LYP / 6-311+G(d)}   \\
\cmidrule{2-3} \cmidrule{4-5}
 &$\lambda_{adia}$ $(nm)$ &$\lambda_{0-0}$ $(nm)$&$\lambda_{adia}$ $(nm)$& $\lambda_{0-0}$ $(nm)$  \\
\midrule
anti-\ce{B18H22}           & 340 [344]& 351 [355]& 347 [350]& 358 [362]       \\
7-F-anti-\ce{B18H21}       & 346 [350]& 357 [362]& 353 [357]& 365 [370]       \\
4-F-anti-\ce{B18H21}       & 353 [356]& 365 [369]& 359 [362]& 373 [375]       \\
4-Br-anti-\ce{B18H21}      & 405 [383]& 419 [396]& 422 [398]& 438 [412]       \\
\bottomrule
\end{tabular*}
\footnotetext{$\lambda_{adia}$ adiabatic wavelength, $\lambda_{0-0}$ origin band wavelength. 
The values in bracket were calculated with CPCM(Hexane) model.}
\end{table}

\section{Conclusion}
In summary, our work delves into the quantum chemical analysis of mono-halogenated borane molecules utilizing DFT and TD-DFT methods. 
We compare the archetypal anti-\ce{B18H22} with hypothetical halogenated derivatives: 7-F-anti-\ce{B18H21}, 4-F-anti-\ce{B18H21}, 
and the recently synthesized 4-Br-anti-\ce{B18H21}. 
Our comprehensive study includes optimizations of ground state and first singlet excited state geometries, vibrational frequency analysis, 
and detailed spectroscopic characterization. 
The fluorinated compounds display distinct features in the IR spectra, while the brominated compound exhibits a unique peak in the Raman spectra. 
UV-Vis analysis highlights changes in electronic properties due to halogenation, and all studied compounds emit visible light, 
indicating their potential for optoelectronic applications. 
Ultimately, using ORCA software with the PBE0/def2-SVPD and B3LYP/6-311+G(d) methods, alongside the CPCM model, 
achieves accuracy comparable to the more computationally demanding CASPT2 method. 
This convergence significantly streamlines the theoretical investigation, particularly for the brominated derivative.

One can substitute other halogens in the used two positions, in order to study the effects of the substituting atom identity and the substitution site on the spectral properties of Borane Laser. 

All studied compounds were found to emit visible light, suggesting their potential for optoelectronic applications. Actually, one can signal two impacts of shifts in absorption peaks, the first on wavelength range which may widen under halogenation, which is crucial for applications like solar cells with broader absorption spectrums improving efficiency, whereas the second relates to changes in the absorption intensity, shown by the damping coefficient, with higher intensity meaning better light-harvesting capabilities, essential for light-emitting devices. As to the shift in emission peaks, it allows for color tuning which is important for display technologies and lighting applications. The best halogenated compound is revealed to be the compound (2) with F-substituted at site 7, because it has the highest absorptivity, in a similar way to the Cl-substituted compound studied previously and which showed a better quantum yield when compared to other halogenated compounds.     

\bmhead{Supplementary information}

\bmhead{Acknowledgements}
N. C. acknowledges support from the CAS PIFI fellowship and from the Humboldt Foundation.


\bibliography{QISP_article_ref}
\newpage

\begin{appendices}
\setlength{\tabcolsep}{1pt} 

\section{Vertical absorption wavelengths and corresponding oscillator strength}\label{secA1}

\begin{table}[htbp]
\caption{Vertical absorption wavelengths and corresponding oscillator strength $f$. Using PBE0 / def2-SVPD}
\label{pbe0}
\begin{tabular*}{\textwidth}{@{\extracolsep\fill}lcccccccc}
\toprule
\multirow{2}{*}{State} & \multicolumn{2}{c}{anti-\ce{B18H22}} & \multicolumn{2}{c}{7-F-anti-\ce{B18H21}} & \multicolumn{2}{c}{4-F-anti-\ce{B18H21}} 
& \multicolumn{2}{c}{4-Br-anti-\ce{B18H21}} \\
\cmidrule{2-9}
 & $\lambda_{VA}$  & $f$ & $\lambda_{VA}$ & $f$ & $\lambda_{VA}$  & $f$ & $\lambda_{VA}$ & $f$ \\
\midrule
\ce{S1} & 307.9 & 0.1448 & 312.3 & 0.1637 & 316.1 & 0.1458 & 354 & 0.0627 \\
\ce{S2} & 263.3 & 0.0000 & 262 & 0.0009 & 272.3 & 0.0533 & 341.1 & 0.0368 \\
\ce{S3} & 258.9 & 0.0564 & 256.1 & 0.0471 & 261.2 & 0.0176 & 294.5 & 0.0809 \\
\ce{S4} & 248.2 & 0.0000 & 248.3 & 0.0027 & 246.8 & 0.0119 & 257.6 & 0.024 \\
\ce{S5} & 240.7 & 0.0275 & 240.8 & 0.0144 & 243.1 & 0.0017 & 251.4 & 0.0065 \\
\ce{S6} & 232.9 & 0.0000 & 234.9 & 0.0006 & 235.7 & 0.0166 & 247.7 & 0.0024 \\
\ce{S7} & 230.4 & 0.0164 & 230.3 & 0.0296 & 233.2 & 0.0012 & 243.8 & 0.0046 \\
\ce{S8} & 229.2 & 0.0000 & 229 & 0.0088 & 229.9 & 0.0215 & 236.5 & 0.0377 \\
\ce{S9} & 226.6 & 0.0346 & 223.4 & 0.0039 & 228.3 & 0.0061 & 234.6 & 0.0071 \\
\ce{S10} & 224.8 & 0.0000 & 221.4 & 0.0002 & 225.8 & 0.0041 & 231.9 & 0.0033 \\
\ce{T1} & 376.9 &  forbidden & 384.9 &  forbidden & 387.1 &  forbidden & 403.3 &  forbidden \\
\ce{T2} & 288.3 &  forbidden & 290.2 &  forbidden & 304.9 &  forbidden & 359.6 &  forbidden \\
\ce{T3} & 286.7 &  forbidden & 284 &  forbidden & 291.7 &  forbidden & 327.9 &  forbidden \\
\ce{T4} & 267 &  forbidden & 264.5 &  forbidden & 262.3 &  forbidden & 277.7 &  forbidden \\
\ce{T5} & 256.2 &  forbidden & 254.8 &  forbidden & 255.9 &  forbidden & 271.5 &  forbidden \\
\ce{T6} & 248.8 &  forbidden & 251.2 &  forbidden & 251.6 &  forbidden & 264.1 &  forbidden \\
\ce{T7} & 243.1 &  forbidden & 245.3 &  forbidden & 249.5 &  forbidden & 256.1 &  forbidden \\
\ce{T8} & 240.2 &  forbidden & 237.7 &  forbidden & 240.3 &  forbidden & 251.7 &  forbidden \\
\ce{T9} & 237.5 &  forbidden & 232 &  forbidden & 237.3 &  forbidden & 241.5 &  forbidden \\
\ce{T10} & 232 &  forbidden & 231.8 &  forbidden & 234.7 &  forbidden & 240.4 &  forbidden \\
\bottomrule
\end{tabular*}
\footnotetext{$\lambda_{VA}$ vertical absorption wavelength, $f$ oscillator strength.}
\end{table}

\begin{table}[htbp]
\caption{Vertical absorption wavelengths and corresponding oscillator strength $f$. Using PBE0 / def2-SVPD / Hexane}
\label{pbe0s}
\begin{tabular*}{\textwidth}{@{\extracolsep\fill}lcccccccc}
\toprule
\multirow{2}{*}{State} & \multicolumn{2}{c}{anti-\ce{B18H22}} & \multicolumn{2}{c}{7-F-anti-\ce{B18H21}} & \multicolumn{2}{c}{4-F-anti-\ce{B18H21}} 
& \multicolumn{2}{c}{4-Br-anti-\ce{B18H21}} \\
\cmidrule{2-9}
 & $\lambda_{VA}$  & $f$ & $\lambda_{VA}$ & $f$ & $\lambda_{VA}$  & $f$ & $\lambda_{VA}$ & $f$ \\
\midrule
\ce{S1} & 309.8 & 0.2014 & 314.5 & 0.2229 & 317.3 & 0.2022 & 341.3 & 0.1189 \\
\ce{S2} & 262.3 & 0.0000 & 261.5 & 0.0015 & 271.1 & 0.0774 & 324.8 & 0.0515 \\
\ce{S3} & 259.1 & 0.0846 & 256.2 & 0.0721 & 260.7 & 0.024 & 291.3 & 0.0829 \\
\ce{S4} & 247.8 & 0.0000 & 247.1 & 0.0032 & 246 & 0.0176 & 256.9 & 0.037 \\
\ce{S5} & 240.2 & 0.0453 & 240.1 & 0.0247 & 242.1 & 0.0034 & 249.9 & 0.0056 \\
\ce{S6} & 232.1 & 0.0000 & 234.3 & 0.0009 & 234.9 & 0.0289 & 246.8 & 0.0073 \\
\ce{S7} & 229.5 & 0.0304 & 230.3 & 0.0489 & 232.6 & 0.0039 & 239.5 & 0.0059 \\
\ce{S8} & 229 & 0.0000 & 228.1 & 0.0068 & 229.6 & 0.0317 & 235.6 & 0.0498 \\
\ce{S9} & 226.5 & 0.0409 & 223 & 0.0052 & 227.6 & 0.0048 & 231.2 & 0.0021 \\
\ce{S10} & 224.3 & 0.0000 & 220.9 & 0.0001 & 225.2 & 0.0052 & 228.6 & 0.0092 \\
\ce{T1} & 375.9 & forbidden & 383.6 & forbidden & 385.3 & forbidden & 395.1 & forbidden \\
\ce{T2} & 287.8 & forbidden & 289.5 & forbidden & 302.7 & forbidden & 343.1 & forbidden \\
\ce{T3} & 286 & forbidden & 283.7 & forbidden & 290.6 & forbidden & 318.2 & forbidden \\
\ce{T4} & 265.4 & forbidden & 262.7 & forbidden & 261 & forbidden & 275.4 & forbidden \\
\ce{T5} & 255.7 & forbidden & 254.1 & forbidden & 255 & forbidden & 268.6 & forbidden \\
\ce{T6} & 248.4 & forbidden & 250.8 & forbidden & 251.1 & forbidden & 262.9 & forbidden \\
\ce{T7} & 242.3 & forbidden & 244.7 & forbidden & 248.4 & forbidden & 253.9 & forbidden \\
\ce{T8} & 239.2 & forbidden & 236.4 & forbidden & 239.4 & forbidden & 249.9 & forbidden \\
\ce{T9} & 236.3 & forbidden & 231.5 & forbidden & 236.2 & forbidden & 240.3 & forbidden \\
\ce{T10} & 231.3 & forbidden & 231.4 & forbidden & 233.7 & forbidden & 237.5 & forbidden \\
\bottomrule
\end{tabular*}
\footnotetext{$\lambda_{VA}$ vertical absorption wavelength, $f$ oscillator strength.}
\end{table}

\begin{table}[htbp]
\caption{Vertical absorption wavelengths and corresponding oscillator strength $f$. Using B3LYP / 6-311+G(d)}
\label{b3lyps}
\begin{tabular*}{\textwidth}{@{\extracolsep\fill}lcccccccc}
\toprule
\multirow{2}{*}{State} & \multicolumn{2}{c}{anti-\ce{B18H22}} & \multicolumn{2}{c}{7-F-anti-\ce{B18H21}} & \multicolumn{2}{c}{4-F-anti-\ce{B18H21}} 
& \multicolumn{2}{c}{4-Br-anti-\ce{B18H21}} \\
\cmidrule{2-9}
 & $\lambda_{VA}$  & $f$ & $\lambda_{VA}$ & $f$ & $\lambda_{VA}$  & $f$ & $\lambda_{VA}$ & $f$ \\
\midrule
\ce{S1}  & 311.2 [313.2] & 0.1415 [0.1962] & 316   [318.2] & 0.1587 [0.2157] & 319.7 [320.9] & 0.1407 [0.1951] & 362.6 [348.7] & 0.0527 [0.1020] \\
\ce{S2}  & 264   [263.2] & 0.0000 [0.0000] & 263.9 [263.4] & 0.0007 [0.0011] & 275.2 [273.9] & 0.0540 [0.0778] & 350.5 [333.6] & 0.0341 [0.0472] \\
\ce{S3}  & 261.4 [261.7] & 0.0569 [0.0849] & 258.9 [259.1] & 0.0469 [0.0715] & 264.1 [263.5] & 0.0138 [0.0190] & 298.8 [296.3] & 0.0861 [0.0928] \\
\ce{S4}  & 249   [248.4] & 0.0000 [0.0000] & 249.4 [248.1] & 0.0020 [0.0023] & 247.2 [246.3] & 0.0098 [0.0138] & 259.8 [259.1] & 0.0255 [0.0396] \\
\ce{S5}  & 241.1 [240.4] & 0.0173 [0.0285] & 241.1 [240.3] & 0.0079 [0.0138] & 243.7 [242.6] & 0.0006 [0.0010] & 253.3 [251.4] & 0.0057 [0.0072] \\
\ce{S6}  & 234.5 [234.1] & 0.0000 [0.0000] & 239.3 [238.7] & 0.0001 [0.0002] & 237.9 [237.3] & 0.0020 [0.0029] & 250.9 [248.1] & 0.0025 [0.0032] \\
\ce{S7}  & 231.1 [230.4] & 0.0000 [0.000]  & 230.7 [230.8] & 0.0266 [0.0409] & 235.4 [234.5] & 0.0130 [0.0258] & 247.4 [244.6] & 0.0031 [0.0045] \\
\ce{S8}  & 230.5 [229.7] & 0.0154 [0.0305] & 228.9 [227.9] & 0.0022 [0.0028] & 230.1 [229.8] & 0.0170 [0.0236] & 241.8 [236.6] & 0.0227 [0.0371] \\
\ce{S9}  & 226.9 [226.6] & 0.0241 [0.0271] & 224.2 [223.9] & 0.0027 [0.0039] & 229.8 [229.1] & 0.0007 [0.0002] & 237.2 [233.4] & 0.0092 [0.0156] \\
\ce{S10} & 225.9 [225.4] & 0.0000 [0.0000] & 221.3 [220.9] & 0.0002 [0.0003] & 226.6 [226  ] & 0.0053 [0.0069] & 231.8 [231.1] & 0.0014 [0.0071] \\
\ce{T1}  & 376.6 [375.8] & - [-]           & 384.9 [383.7] & -      [-     ] & 386.9 [385.1] & -      [-     ] & 405.8 [396.8] & -      [-     ] \\
\ce{T2}  & 288   [287.5] & - [-]           & 290.5 [289.8] & -      [-     ] & 304.6 [302.4] & -      [-     ] & 366.4 [349  ] & -      [-     ] \\
\ce{T3}  & 286.2 [285.6] & - [-]           & 284   [283.7] & -      [-     ] & 292.1 [290.9] & -      [-     ] & 332.6 [323  ] & -      [-     ] \\
\ce{T4}  & 265.3 [263.7] & - [-]           & 263.1 [261.3] & -      [-     ] & 260.9 [259.5] & -      [-     ] & 278.7 [276.8] & -      [-     ] \\
\ce{T5}  & 253.9 [253.3] & - [-]           & 253.8 [253.1] & -      [-     ] & 255.6 [254.7] & -      [-     ] & 272.9 [270.1] & -      [-     ] \\
\ce{T6}  & 249.6 [249.3] & - [-]           & 252.2 [251.5] & -      [-     ] & 251.6 [250.6] & -      [-     ] & 262.3 [261  ] & -      [-     ] \\
\ce{T7}  & 242.2 [241.5] & - [-]           & 244.6 [244  ] & -      [-     ] & 248.3 [247.6] & -      [-     ] & 259.5 [255.8] & -      [-     ] \\
\ce{T8}  & 238.1 [237.2] & - [-]           & 236   [234.9] & -      [-     ] & 239.8 [238.9] & -      [-     ] & 250.6 [249.5] & -      [-     ] \\
\ce{T9}  & 236   [234.8] & - [-]           & 234.7 [234.3] & -      [-     ] & 237.5 [236.6] & -      [-     ] & 246.5 [240.1] & -      [-     ] \\
\ce{T10} & 233.1 [232.4] & - [-]           & 232.7 [232.3] & -      [-     ] & 235   [234  ] & -      [-     ] & 242.4 [238.4] & -      [-     ] \\
\bottomrule
\end{tabular*}
\footnotetext{$\lambda_{VA}$ vertical absorption wavelength, $f$ oscillator strength.}
\end{table}

\end{appendices}

\end{document}